\documentclass[aps,prd,superscriptaddress,twocolumn,
]{revtex4-1}

\usepackage{amsmath,amssymb,amsfonts,bm,graphicx}
\usepackage{mathtools}
\usepackage[normalem]{ulem}
\usepackage[usenames, dvipsnames]{xcolor}
\usepackage{cancel}

\usepackage{hyperref}
\hypersetup{colorlinks=true, linkcolor=blue, citecolor=blue, urlcolor=blue}

\usepackage[T1]{fontenc}

\newcommand\blue[1]{{\color{black}#1}}

\begin{document}
\title{
Stability and electronic properties of LaNiO$_2$/SrTiO$_3$ heterostructures
}

\author{F. Bernardini}
%\email{fabio.bernardini@dsf.unica.it}
\affiliation{Dipartimento di Fisica, Universit\`a di Cagliari, IT-09042 Monserrato, Italy}

\author{A. Cano}
\affiliation{Institut N\'eel, CNRS \& UGA, 38042 Grenoble, France}
%\email{andres.cano@cnrs.fr}
\date{\today}

\begin{abstract}
Infinite-layer nickelate thin films materialize an intriguing new platform for high-temperature unconventional superconductivity, with LaNiO$_2$/SrTiO$_3$ as reference setup. We discuss the relative stability of the elementary interfaces of this system and determine the corresponding electronic band structure. We find substantial changes compared to the bulk, in particular in relation to the 5$d$ orbital contributions to the low-energy physics which can be totally replaced by purely Ni-3$d$ flat bands. The $d^9$ configuration characteristic of cuprates can thus be supplemented by an extra interfacial ingredient destabilizing the normal non-superconducting state 
in these heterostructures. 
\end{abstract}

\maketitle

Infinite-layer nickelates have long been proposed as intriguing analogues to high-$T_c$ superconducting cuprates, thus sustaining a rather fundamental research on these systems over the years \cite{hayward99,anisimov99,pickett04,crespin05,botana17,botana17a}. This is motivated by the nominal 3$d^9$ electronic configuration featured by the Ni and Cu atoms in both these systems, which is however accompanied by a self-doping effect due to rare-earth 5$d$ states in the case of \blue{$R$NiO$_2$} nickelates \blue{($R=$ rare-earth atom)}. As a result of such an activity, superconductivity has recently been reported in Sr-doped NdNiO$_2$/SrTiO$_3$ thin films \cite{li-nature19,li2020superconducting}. 
This finding represents an important breakthrough that, however, seems to be remarkably difficult to reproduce. In fact, the reported $T_c$'s vary quite substantially from sample to sample ---even if they are nominally equivalent--- and no definitive result is given for the LaNiO$_2$/SrTiO$_3$ case \cite{li-nature19}. 
\blue{The latter has been tentatively ascribed to the possible incorporation of topotactic hydrogens that, in practice, would lead to a different $R$NiO$_2$H phase being probed in the experiments \cite{si2019topotactic}.} 
Beyond that, no superconductivity has been found in epitaxial thin films directly grown on SrTiO$_3$ and alternative substrates and, very importantly, in bulk samples so far \cite{zhou2019absence,li2019absence}.  
This state of affairs calls for the investigation of the corresponding interface itself since it can play a non-trivial role in the reported superconductivity. 

\begin{figure*}[t!]
\includegraphics[width=.85\textwidth]{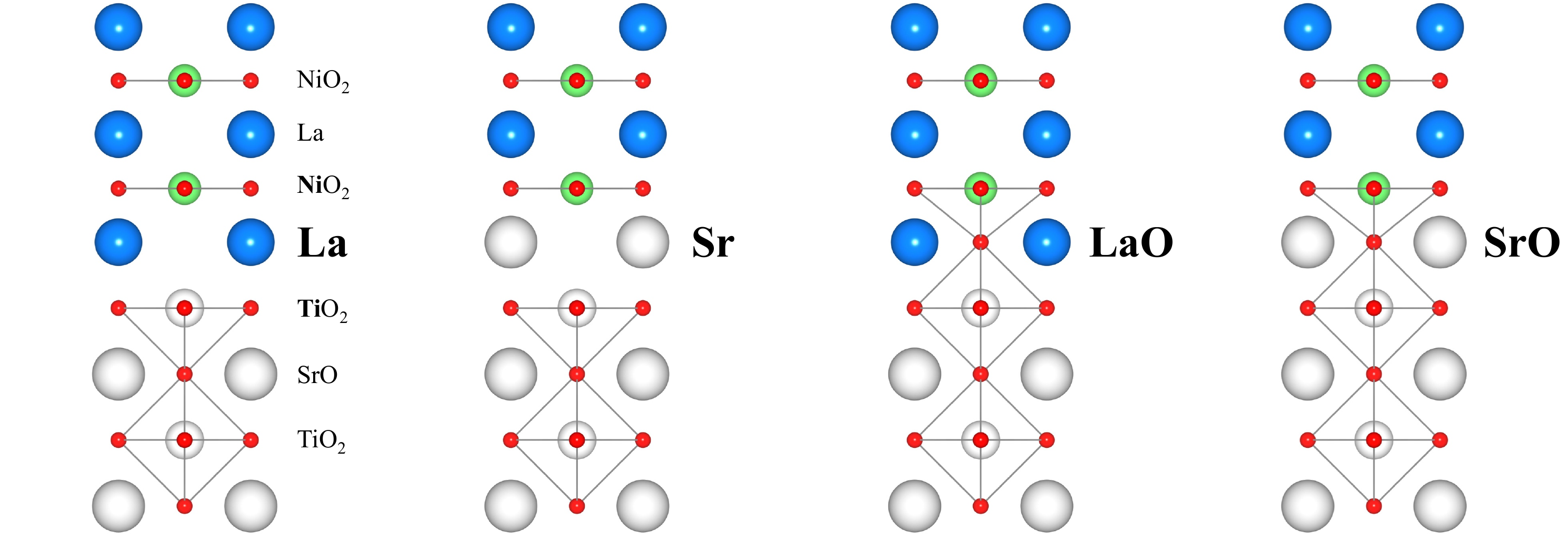}
  \caption{Ball-and-stick model of the elementary LaNiO$_2$/SrTiO$_3$ interfaces. The boundary layer can be made of either La, Sr, LaO, or SrO planes as indicated in the figure. The interfacial atoms indicated with bold symbols are denoted by the $i$ subscript in the main text.}  
\label{fig:interfaces}
\end{figure*}

Here, we investigate theoretically the fundamental properties of the reference LaNiO$_2$/SrTiO$_3$ interface by means of density functional theory (DFT) calculations. This system is expected to be representative for the \blue{infinite-layer} nickelate series without requiring any ambiguous treatment of 4$f$ electrons \cite{anisimov99,pickett04,botana19,sakakibara19,wu19,bernardini19}. In addition, the DFT band structure near the Fermi level has proven to be rather robust with respect to many-body correlation effects in bulk LaNiO$_2$ \cite{olevano20}. 
We model the interface by a symmetric supercell with two identical interfacial layers among the four elementary possibilities illustrated in Fig. \ref{fig:interfaces}. 
Further, one extra LaNiO$_2$ layer is sandwiched between these interfaces, so that our model also has a connection with trilayer bulk systems such as $R_4$Ni$_3$O$_8$ \cite{botana20trilayer}.
We first determine the most stable interfacial configuration according to the corresponding thermodynamics. 
As we show below, the fragile stability of the infinite-layer bulk nickelates manifests also through the marked differences in the relative energetics of the heterostructure. Thus, the growth process can be expected to have a non-negligible impact on the eventual atomic configuration that is realized experimentally. 
In addition, we study the specific features that emerge locally in the band structure of the system as a function of 
its interfacial atomic configuration. We thus examine the most distinctive features of the superconducting nickelates, with a focus on the self-doping effect of the 5$d$ states.
In fact, we find drastic changes compared to the bulk. 
While the Ni-3$d_{x^2-y^2}$ low-energy features are robust against interfacial effects, the metallic character provided by these 5$d$
states in the bulk can be replaced either by Ti-3$d$ contributions or directly by Ni-3$d_{z^2}$ flat bands in the ultrathin limit. This modifies qualitatively the initial picture for the development of strong correlations and the eventual Cooper pairing, which then might display a distinct interfacial nature in infinite-layer nickelates of this class.    

\section{Computational Methods \label{methods}} 
We performed density functional theory (DFT) calculations to investigate the LaNiO$_2$/SrTiO$_3$ heterostructure. We considered a tetragonal supercell with 3 Ni and 3 Ti atoms and two identical boundary layers, simulating the epitaxial LaNiO$_2$ by imposing the calculated $a=b=3.94$~{\AA} lattice parameters of bulk SrTiO$_3$. The $c$ parameter and the internal atomic positions, in their turn, were optimized using a constrained variable-cell dynamics. In this way, we avoid the presence of the residual stresses that might falsify the relative stability of the different interfacial configurations. We used the VASP code~\cite{VASP} for these \blue{spinless} calculations with the PBE~\cite{PBE} exchange-correlation functional and PAW pseudo-potentials~\cite{PAW}. We employed a plane-wave cutoff of 540 eV and a Monkhorst-Pack $6\times6\times2$ $k$-mesh with a 0.2 eV Gaussian smearing, and treated the Sr-4$s$4$p$, Ti-3$s$3$p$ and Ni-3$p$ electrons as valence electrons. The convergence criteria were 1 meV on the total energy, 0.001 \AA ~on the $c$ parameter, and 0.01 eV/\AA ~ on the residual forces. 
The structural parameters of the supercells are summarized in Table~\ref{tab:SL-struct}. The corresponding chemical potentials are determined from analogous total energy calculations of equilibrium bulk structures using the same plane-wave cutoff and smearing, and a $k$-mesh equivalent to a cubic 6$\times$6$\times$6 one whenever possible. The results are summarized in Table~\ref{tab:bulks} in the Appendix.

The electronic band structure was further computed using the full-potential linear augmented plane-wave (FLAPW) method as implemented in the {\sc{WIEN2k}} package \cite{Wien2k}, with the LDA exchange-correlation functional \cite{LDA,note-dft}. 
We performed spinless calculations with muffin-tin radii of 2.5, 2.1, 2.0 and 1.5 Bohr for the La (Sr), Ni, Ti and O atoms respectively and a plane-wave cutoff $R_{\rm MT}K_{\rm max}=7.0$. The integration over the Brillouin zone was performed using a 13$\times$13$\times$2 $k$-mesh for the self-consistent calculations, while a denser 36$\times$36$\times$6 $k$-mesh was used for the Fermi surface.

\begin{table}[t!]
\begin{tabular}{l c c c c}
\hline \hline
                   & La       &     Sr & LaO    & SrO    \\    \hline
$c$ (\AA)          &  21.82   &  21.89 & 22.50  &  22.70 \\
Ni$_i$-O    (\AA)  &  1.982   &  1.970 & 1.977  &  1.994 \\
O$_i$-Ni$_i$(\AA)  &  -       &  -     & 2.196  &  1.966 \\
$\angle$ O$_i$-Ni$_i$-O ($^\circ$) &  -       &   -  & 85.07 &  98.95 \\ 
$E_{\rm tot}$ (eV)  & $-$201.07 & $-$192.92  & $-$218.69 & $-$209.13 \\
\hline \hline
\end{tabular}
\caption{Structural parameters and total energy of the supercells used to study the four elementary LaNiO$_2$/SrTiO$_3$ interfaces sketched in Fig. \ref{fig:interfaces}. Ni$_i$ denotes the first nickel next to the interfacial layer and O$_i$ the interfacial oxygen (bold symbols in Fig. \ref{fig:interfaces}). The reported values correspond to the overall $c$ parameter, O$_{i}$-Ni$_{i}$ distance, Ni$_{i}$-O distance, and O$_i$-Ni$_{i}$-O bond angle.}
\label{tab:SL-struct}
\end{table}

\section{Structure of the interface}

We first address the question of the actual structure of the LaNiO$_2$/SrTiO$_3$ interface. 
We assume the ideal case in which the interface is not modified by e.g. oxygen vacancies or topotactic hydrogens \cite{si2019topotactic}, and such that no interface reconstruction takes place. The latter indeed seems to be the case according to the scanning transmission electron microscopy (STEM) images reported in \cite{lee2020aspects}.
In fact, in (001) oriented SrTiO$_3$ substrates, ideal SrO- and TiO$_2$-terminated surfaces are the most stable configurations according to first-principles calculations \cite{Heifets} and they regularly are realized experimentally \cite{Szot}.

In order to determine the relative formation energy of the elementary LaNiO$_2$/SrTiO$_3$ interfaces (see Fig. \ref{fig:interfaces}) we follow the well-established 
thermodynamic approach described in  \cite{Qian1988,Chetty1992,vdWalle1993}. 
Thus we define the interface energy as
\begin{equation}
     E_{\rm interface}={1\over 2}\Big( E_{\rm tot}
     -\sum_\text{X=La, Sr, Ni, Ti, O}
     \mu_{\rm X}N_{\rm X}\Big),
     \label{Einterface}\end{equation}
where $E_{\rm tot}$ is the total energy of the corresponding supercell, $\mu_X$ is the chemical potential of the X element, and $N_{\rm X}$ is the number of X atoms in the supercell. 
In this way, the energy difference between the La and LaO configurations for example can be written as
\begin{align}
    \Delta E_\text{interface}^{\rm La - LaO}
    &= \frac{1}{2} \big( E_{\rm tot}^{\rm (La)}-E_{\rm tot}^{\rm (LaO)}\big)
    +\mu_{\rm O} \nonumber \\
    &=\frac{1}{2} \big( E_{\rm tot}^{\rm (La)}-E_{\rm tot}^{\rm (LaO)}+E_{\rm tot}^{\rm O_2}\big)+\Delta\mu_{\rm O}, \label{SrO-Sr}    
\end{align}
and likewise for the rest. 
Here and hereafter $\Delta\mu_{\rm X}\leq 0$ denotes the chemical potential of X relative to its value in the most stable elementary phase of X (that is, $2 \Delta\mu_{\rm O} =2 \mu_{\rm O} - E_{\rm tot}^{\rm O_2}$ in the case of X = O). Thus, the relative stability of the different configurations can be nailed down from these chemical potentials, which can be estimated as detailed in Appendix \ref{appendix}. The results, summarized in Fig. \ref{fig:LaNiOnSTO}, are as follows.

The two-step synthesis of the samples performed in \cite{li-nature19} starts with the epitaxial growth of the perovskite heterostructure ---that is, LaNiO$_3$/SrTiO$_3$ in our case. At this stage, the interfacial configuration of the system restricts to either LaO or SrO. In Fig. \ref{fig:LaNiOnSTO}, the horizontal thick line in blue represents the boundary between these two possibilities. Above (below) this line, the LaO (SrO) configuration minimizes the overall energy of the system. However, in thermodynamic equilibrium, the chemical potentials have to be inside the yellow region. As a result, the two configurations can be obtained under oxygen-rich conditions while only the LaO one is compatible with the oxygen-poor case ($\Delta \mu_\text{O} > $ and $<-1.9$~eV respectively). 

\begin{figure}[t!]
\includegraphics[width=.4\textwidth]{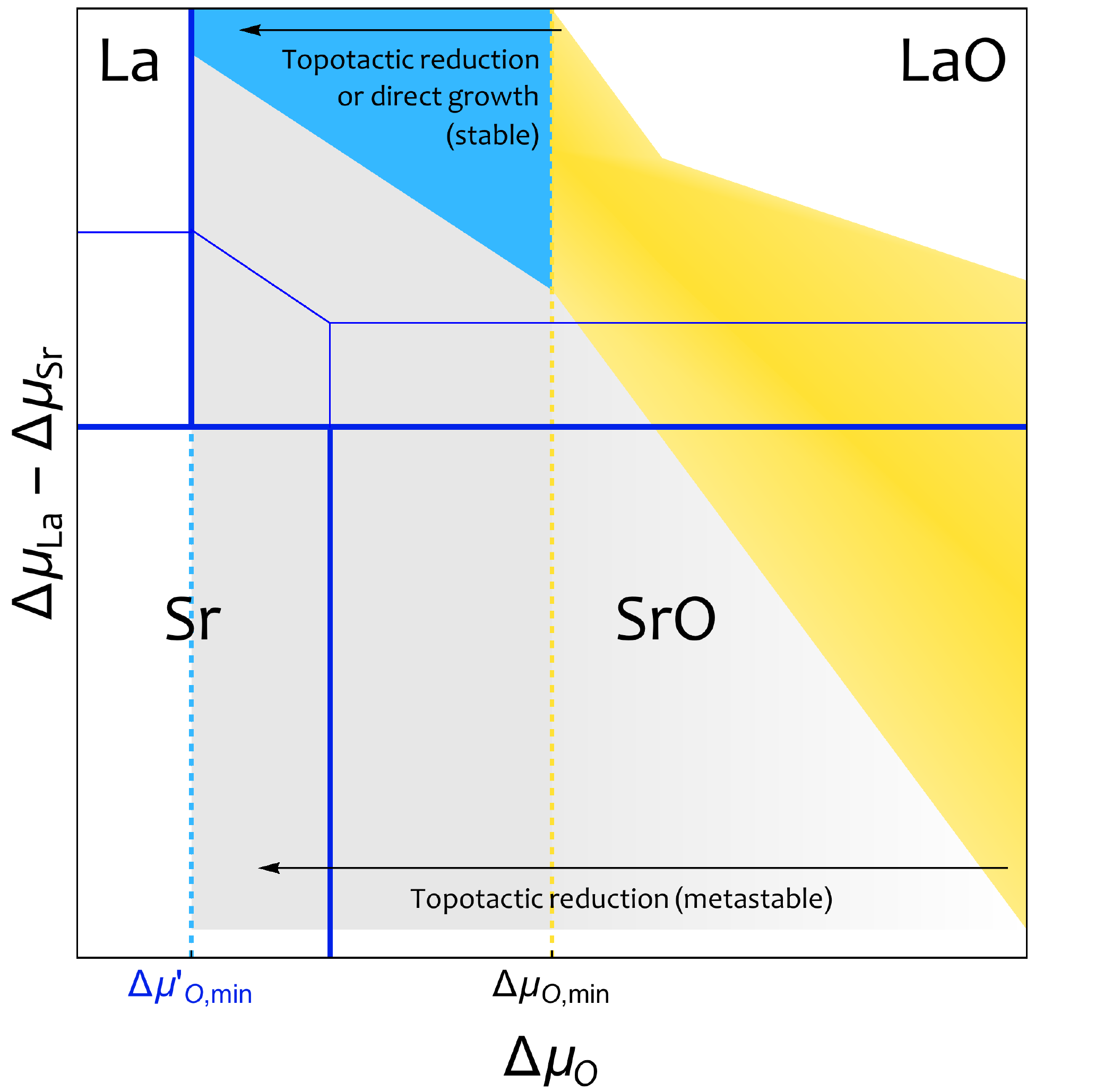}
\caption{
\label{fig:LaNiOnSTO} Calculated phase diagram for the interfacial configuration of the LaNiO$_n$/SrTiO$_3$ heterostructure ($n=2,3$). LaO, SrO, La, and Sr label different regions according to the relative stability of the corresponding interfacial configurations (the thin lines apply for the direct growth of $n=2$). The yellow area indicates the thermodynamically allowed (equilibrium) values of the chemical potentials for $n=3$, while the blue does for $n = 2$ (as obtained by both \blue{topotactic} reduction and direct growth). The dashed vertical lines indicate the minimum allowed value of the oxygen chemical potential before and after the topotactic reduction (the dashed blue line also applies for the direct growth for $n=2$). Both LaO and SrO configurations can be realized in the LaNiO$_3$/SrTiO$_3$ perovskite heterostructure, while only LaO becomes compatible with the stability of the bulk infinite-layer phase in LaNiO$_2$/SrTiO$_3$. The interfacial SrO, Sr (and La) configurations, in contrast, imply the thermodynamic metastability of the system in the latter heterostructure.  
}    

\end{figure}

The subsequent topotactic reduction removes oxygens from the nickelate, which in principle enables the additional Sr and La interfacial configurations from the previous ones. This introduces the new LaO-La and SrO-Sr boundaries indicated by the vertical solid lines in Fig. \ref{fig:LaNiOnSTO}. In terms of the chemical potentials, the reduction of the system implies the lowering of the minimum allowed value of the oxygen chemical potential $\Delta \mu_\text{O,min}$ (from $-2.5$~eV to $-4.4$~eV as indicated by the dashed vertical lines in Fig. \ref{fig:LaNiOnSTO}). The La configuration, however, never minimizes the energy above $\Delta \mu_\text{O,min}'$ and therefore will not be realized unless extra oxygen-vacancies be present. The other three configurations are still compatible with the ideal reduction. However, the infinite-layer LaNiO$_2$ phase of the bulk nickelate is stable only within the blue region. Consequently, the SrO and Sr configurations can appear but as metastable states of the system, while the LaO configuration is the only one that is in thermodynamic equilibrium with the bulk LaNiO$_2$. 
The analysis of the direct growth of the LaNiO$_2$/SrTiO$_3$ heterostructure  qualitatively yields the same result (with the LaO-SrO, LaO-Sr and La-Sr boundaries replaced by the thin blue lines in Fig. \ref{fig:LaNiOnSTO}).    
These findings are confirmed by the STEM images reported in \cite{lee2020aspects}, in the sense that oxygen atoms are visible at the interface.

\begin{figure*}[t!]
\includegraphics[width=1\textwidth]{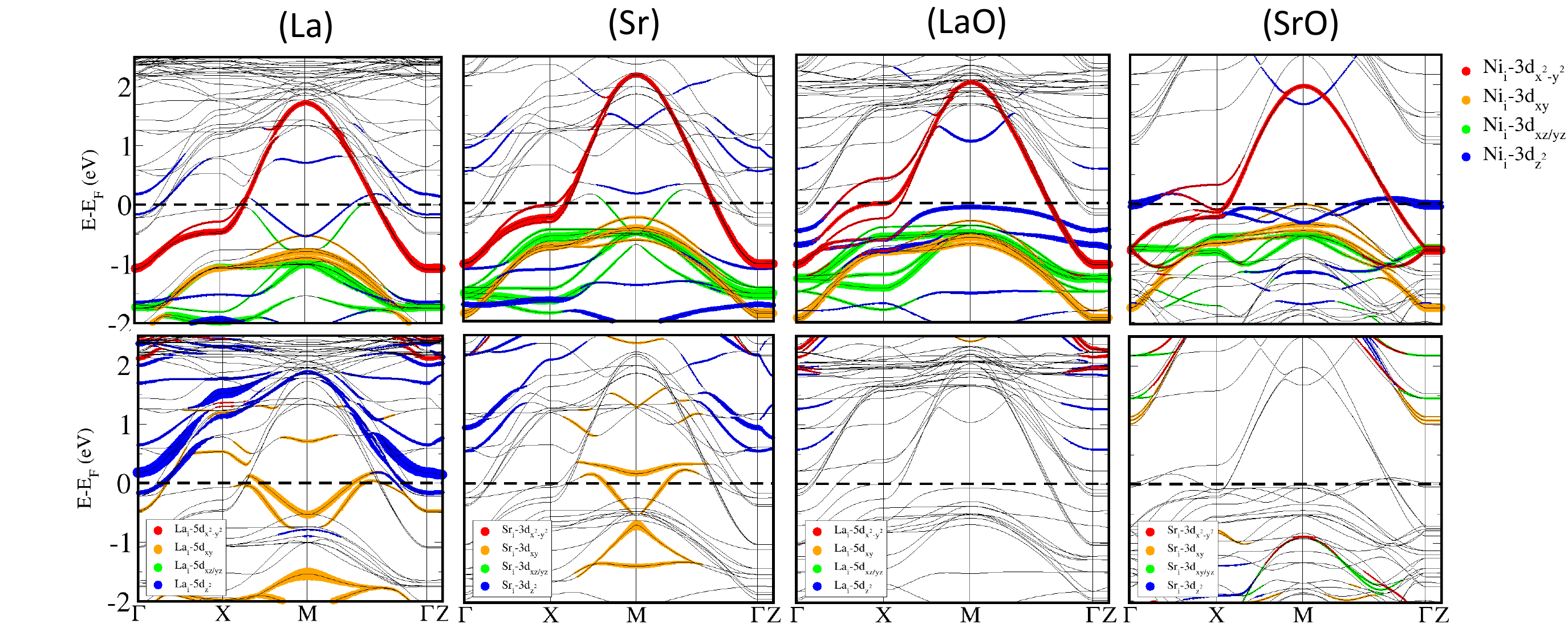}
  \caption{
  Band structure of the LaNiO$_2$/SrTiO$_3$ symmetric supercells for different interfacial configurations. The line thickness and the different colors highlight the interfacial contributions according to their atomic-orbital content.}   
\label{fig:SLbands}
\end{figure*}

\section{Electronic band structure}

Next, we discuss the electronic band structure of the LaNiO$_2$/SrTiO$_3$ heterostructure as a function of the \blue{interfacial configuration}. 
In our calculations 
we considered supercells containing 2 equivalent interfacial Ni$_i$ atoms + 1 ``bulk'' Ni atoms (surrounded by 1~O$_i$ + 4~O and 4~O atoms respectively, see Fig.~\ref{fig:interfaces}). 
Thus, our results apply to the ultrathin limit of the infinite-layer nickelates and have rather direct connection to the trilayer bulk systems $R_4$Ni$_3$O$_8$ recently discussed in \cite{botana20trilayer}. \blue{Here we focus on the results obtained at the LDA level, which are shown to be essentially the same in LDA + $U$ in Appendix \ref{a:LDA+U}.}

It is instructive to start with the La and Sr configurations, even if these are the most unlikely interfaces according to the overall thermodynamics discussed in the previous section. Fig.~\ref{fig:SLbands} shows the calculated dispersion across the two-dimensional Brillouin zone of the system and perpendicular to it.The La case could be anticipated to be rather similar to the bulk since the local environment of the interfacial Ni$_i$ is essentially the same. This is confirmed in Fig. \ref{fig:SLbands} where, in addition to the Ni$_i$-3$d_{x^2-y^2}$ bands crossing the Fermi level, we also observe the self-doping effect due to the La$_i$-5$d_{z^2}$ ones. These are the main features of the bulk, which are thus preserved at the interface in such a La configuration. The main change is observed at the $M$ point, where two additional bands can be seen near the Fermi level. Their Ni$_i$ orbital character is 3$d_{z^2}$ and 3$d_{xz/yz}$ along the $X$-$M$-$\Gamma$ line. 
These features result from the mixing with the La$_i$-5$d_{xy}$ bands, which now provide an important extra contribution to the La-5$d_{z^2}$ electron self-doping at this interface.

In the case of the Sr interface we observe essentially the same features. The main difference now is the slight overall shift upwards of the bands and, in particular, of their Ni-3$d_{z^2}$ character. This is totally in tune with the interfacial hole-doping that results from the La $\to$ Sr local substitution at the interface. This substitution, in addition, has a striking impact on the electron-self doping effect at the interface, which is now surprisingly provided by Ti$_i$-3$d$ states at $\Gamma$ (and not by Sr$_i$-3$d$ ones). Thus, the nature of the metallic character of the system is locally modified by the interfacial configuration.

In the LaO case, more substantial changes are also obtained in the Ni$_i$ bands. First of all, the shift upwards of the originally Ni-3$d_{z^2}$ bands is more pronounced. The mixing with the Ni-3$d_{x^2-y^2}$ ones is thus enhanced and this is accompanied with an important splitting of the Ni and Ni$_i$ bands along the $\Gamma$-$X$-$M$ path about the Fermi level. These features are in fact associated to avoided band crossings that are best exemplified in bulk fluoro-nickelates counterparts \cite{bernardini19b}. The crossing of these bands with the Fermi level is still dominated by the Ni$_i$-3$d_{x^2-y^2}$ orbitals, even if there is a rather flat Ni$_i$-3$d_{z^2}$ band right below $E_F$ at the $M$ point. Beyond that, the bands intersecting the Fermi level at $\Gamma$ have a Ti$_i$-3$d$ character. Again this is in striking contrast to the bulk and can be associated to the interfacial LaO layer itself.

In the case of SrO, the shift upwards of the Ni$_i$-3$d_{z^2}$ bands is even more pronounced and the eventual situation is such that only the Ni bands intersect the Fermi level. They now have both 3$d_{x^2-y^2}$ and 3$d_{z^2}$ orbital contributions. 
The new bands intersecting the Fermi level are remarkably flat, which is compatible with the formation of a genuine interfacial band localized at the SrO plane. Importantly, there is no electron self-doping effect due to neither Ti-3$d$ nor La-5$d$ contributions in this case. The absence of these contributions is an important qualitative difference compared to the previous configurations and the bulk. 

We can gain further insight by discussing these changes in relation to the local atomic structure of the LaO and SrO interfacial layers. In these configurations,
the interfacial Ni$_i$ is surrounded by an extra oxygen O$_i$ compared to the bulk Ni. This oxygen will tend to take one electron from Ti$_i$ and the other from Ni$_i$. Thus, the Ni$_{i}$-O$_{i}$ bond will have an important contribution from the Ni-3$d_{z^2}$ orbitals. Taking into account the rather asymmetric arrangement of the Ni$_i$ environment, this can be expected to yield both bondings and backbondings. In fact, in both LaO and SrO configurations, the flat Ni$_i$-3$d_{z^2}$ band can be associated to such bondings while the empty parabolic band bent upwards with Ni$_i$-3$d_{z^2}$ character at $M$ can be associated to the backbondings. The Ni$_{i}$ will thus tend to feature a nominal Ni$_i^{2+}$ oxidation state in the LaO case (where it is surrounded by 8 La) and Ni$_i^{2.5+}$ in SrO (where it has 4 La and 4 Sr nearby, the latter provinding an extra 0.5 hole doping).  The extra electrons given by the Ni$_i$ will come from the 3$d_{x^2-y^2}$ orbitals, as usual, and from the 3$d_{z^2}$ ones via the interfacial oxygen O$_i$. This explains the shift upwards of the Ni$_i$-3$d_{z^2}$ bands up to the Fermi level. At the same time, this naturally yields a reduced splitting between Ni-3$d$ and O-2$p$ levels (i.e. charge-transfer energies) compared to the bulk. This reduction is such that even a non-negligible O$_i$-2$p$ contribution to density of states eventually emerges at the Fermi energy in the SrO case. Thus, in some sense, these interfacial configurations locally bridge the `charge-transfer vs Mott insulator' gap between bulk cuprates and nickelates \cite{botana19,jiang19}. 

\begin{figure*}[t!]
\includegraphics[width=.85\textwidth]{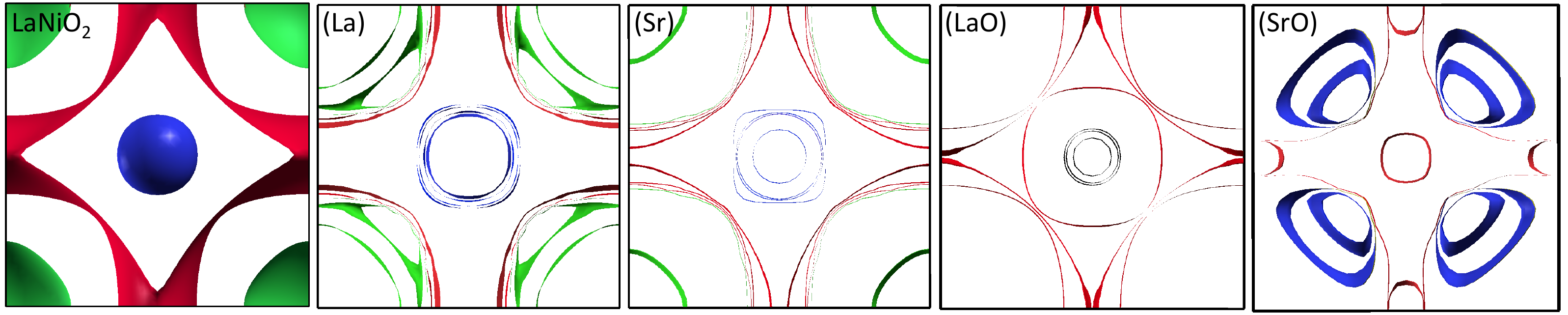}
\caption{Top view of the Fermi surface of bulk LaNiO$_2$ and of the LaNiO$_2$/SrTiO$_3$ heterostructure for the different elementary configurations of the interfacial layer (colors just help to better distinguish the different portions of the Fermi surface). 
}  
\label{fig:SL-FS}
\end{figure*}

We further note that there is an intriguing connection between the reconstructed band structure of the LaNiO$_2$/SrTiO$_3$ heterostructure and that of the trilayer $R_4$Ni$_3$O$_8$ bulk systems \cite{botana20trilayer}. Specifically, the self-doping effect is also absent in these trilayer nickelates where the 5$d$ states are similarly pushed up above the Fermi level. In addition, there is a flattening of the Ni-3$d_{z^2}$ bands ---which additionally shift upwards towards the Fermi level--- together with a reduction of the charge transfer energy. Compared to the parent bulk phase, these changes have been shown to yield a stronger superconducting instability  \cite{botana20trilayer}.

The Fermi surface of the LaNiO$_2$/SrTiO$_3$ heterostructure in the ultrathin limit is shown in Fig.~\ref{fig:SL-FS}. The La, Sr, and LaO interfacial configurations yield a sort of direct 2D version of the bulk LaNiO$_2$. Its self-doping, however, has a different nature. The initial electron pockets at $\Gamma$ (La-5$d$ in the La case as in the bulk) are dominated by Ti-3$d$ orbitals in Sr and LaO, and further disappear in SrO. The La and Sr \blue{interfaces} produce additional pockets at $M$ due to Ni-3$d$ contributions mixed with La-5$d$ and Sr-3$d$ respectively. Most importantly, in the SrO case the initial La-5$d$ pockets at $\Gamma$ disappear and there appears a series of extra pockets along the $\Gamma$-$M$ path. These new pockets are associated to a flat Ni-3$d_{z^2}$ band, which is the most distinct feature of the SrO interfacial configuration. 
\blue{We note that these results are totally compatible with Hall-effect data \cite{li-nature19,hwang20-phasediagram,ariando20-phasediagram} in the sense that they show that the multiband character of the system survives at the interface.}

\section{Conclusions}

Infinite-layer nickelate thin films offer an intriguing new platform for unconventional high-temperature superconductivity, with important analogies and differences with respect to the classic case of cuprates. Our work clarifies the specific fundamental features that emerge in the ultrathin limit of these systems and their connection to interfacial effects. 
To illustrate these features, we have performed a detailed analysis of the reference LaNiO$_2$/SrTiO$_3$ heterostructure. The fragile stability of the infinite-layer nickelates in the bulk is found to have a peculiar impact on the energetics of the elementary interfaces. Thus, while the direct growth of epitaxial films is expected to yield LaO as the most stable interfacial configuration, their two-step synthesis via perovskite precursors is found to be compatible with both SrO and Sr interfacial layers too. This has important consequences for the overall electronic band structure that is eventually realized in these films. 
These can be linked to the local environment of the interfacial Ni atoms. 
The most ``bulk-like'' boundary layer corresponds to the La interfacial configuration, which in fact preserves the main band-structure features of the bulk \cite{note-pentcheva}. The interfacial La $\to$ Sr replacement, however, produces a striking change in the nature of the metallic character of the system, as the extra La-5$d$ contribution becomes Ti-3$d$ at the interface. Thus, the Kondo physics that emerges due to the coupling between the La-5$d$ and Ni-3$d$ electrons in the bulk \cite{zhang-prb20-kondo,zhang19,hepting19} is lost at the interface. The LaO configuration, in addition, implies an enhanced mix of the Ni-3$d_{x^2-y^2}$ and Ni-3$d_{z^2}$ orbitals near the Fermi level. While this has been argued to be detrimental for superconductivity in cuprates \cite{sakakibara12}, the interfacial Ni-3$d_{z^2}$ bands display a remarkable flattening that, by analogy to twisted bilayer graphene and graphite interfaces for example \cite{volovik11,volovik2018graphite}, could result into an additional superconducting instability.     
The SrO interface gives rise to the most dramatic changes. In this case the low-energy physics in the ultrathin limit is in fact entirely determined by Ni-3$d$ electrons, with the original Ni-3$d_{x^2-y^2}$ features now supplemented by the interfacial Ni-3$d_{z^2}$ flat bands exclusively. This changes qualitatively the picture for the subsequent emergence of superconductivity. In fact, trilayer nickelates sharing similar features in the bulk have recently been predicted to host a stronger superconducting instability than their parent infinite-layer phase \cite{botana20trilayer}. Beyond that, the presence of flat bands represents an additional key ingredient that needs to be taken into consideration. 
This is essential in relation to the role of electronic correlations which, even remaining moderate in the bulk \cite{olevano20}, could dominate the low-energy physics at the interface.
Besides, these flat bands can be the source of additional orders that can either cooperate or compete with superconductivity.
Thus, our findings are expected to motivate new perspectives for further theoretical and experimental work on infinite-layer nickelates.

\vspace{1em}
\noindent{\bf Acknowledgments.} The authors are grateful to V. Olevano and X. Blase for insightful discussions. F.B. acknowledges the Visiting Scientist Program of the Centre de Physique Theorique de Grenoble-Alpes (CPTGA) for financial support.

\bibliography{LaNiO2SrTiO3.bib}

\begin{thebibliography}{41}
\bibitem [{not({\natexlab{a}})}]{note-dft}%
  \BibitemOpen
  \href@noop {} {} ({\natexlab{a}}),\ \bibinfo {note} {{ By combining VASP and
  WIEN2k, the relaxed structures and the corresponding band structures are
  obtained with the best accuracy each. While the use of PBE is important for
  the structural optimization, LDA and PBE are equivalent for band structure
  calculations (where the most accurate treatment of localized $d$ and $f$
  orbitals is obtained with WIEN2k).}}\BibitemShut {Stop}%
\end{thebibliography}

\onecolumngrid

%\newpage 

\appendix

\section{Relative stability of the interfacial configurations 
\label{appendix}}

Here, we detail the relative stability of the different interfacial configurations of the relevant heterostructures and the estimate of the chemical potentials. 
We consider both two step-process followed in \cite{li-nature19} and the direct formation of the heterostructure as in \cite{zhou2019absence}. 

\subsection{LaNiO$_3$/SrTiO$_3$ heterostructure}

The elementary interfaces of the perovskite LaNiO$_3$/SrTiO$_3$ heterostructure reduce to either LaO or SrO configurations. Their energy difference can be written as
\begin{align}
    \Delta E_\text{interface}^{\rm SrO - LaO}
    &=\frac{1}{2} \left( E_{\rm tot}^{\rm (SrO)}-E_{\rm tot}^{\rm (LaO)}\right)+E_{\rm tot}^{\rm La}-E_{\rm tot}^{\rm Sr}
    %\nonumber \\&\quad 
    +\Delta\mu_{\rm La}-\Delta\mu_{\rm Sr}, \label{SrO-LaO} 
\end{align}
where the total energies are associated to the corresponding supercells. According to Table \ref{tab:bulks}, we find that the LaO interface will be energetically favored over the SrO one if $\Delta\mu_{\rm La}-\Delta\mu_{\rm Sr}>-2.35$ eV and {\it vice versa}. This corresponds to the horizontal thick line in blue in Fig. \ref{fig:LaNiOnSTO}.

The actual interval of allowed values of the difference $\Delta\mu_{\rm La}-\Delta\mu_{\rm Sr}$ is determined by the minimum and maximum values of the chemical potentials $\Delta\mu_{\rm La}$ and $\Delta\mu_{\rm Sr}$ that are compatible with the overall thermodynamics in equilibrium. These are related to $\Delta\mu_{\rm O}$ as follows.

The existence of bulk LaNiO$_3$ requires
\begin{eqnarray}
\min \Delta\mu_{\rm La} &=& \Delta E_f^{\rm LaNiO_3} 
%-\cancelto{0}{\Delta\mu^{\rm max}_{\rm Ni}}
-3\Delta\mu_{\rm O},\label{eq:La-min-3}\\
\max \Delta\mu_{\rm La} &=& \Delta E_f^{\rm LaNiO_3} -
        \min \Delta\mu_{\rm Ni}-3\Delta\mu_{\rm O}.\label{eq:La-max-3}
\end{eqnarray}
Here we took into account that $\max {\Delta\mu_{\rm Ni}}=0$ in Eq.~\eqref{eq:La-min-3}. On the other hand, $\min \Delta\mu_{\rm Ni}$ must be such that LaNiO$_3$ does not decompose into La$_2$O$_3$, free O and metallic Ni
($E_f^{\rm LaNiO_3} < \frac{1}{2} \Delta E_f^{\rm La2O_3} +\frac{3}{2}\Delta\mu_{\rm O}
    +\Delta\mu_{\rm Ni}$) so that
\begin{equation}
    \min \Delta\mu_{\rm Ni} = 
    E_f^{\rm LaNiO_3} - \frac{1}{2} \Delta E_f^{\rm La2O_3} -\frac{3}{2}\Delta\mu_{\rm O}.
\end{equation}

The minimum value of $\Delta\mu_{\rm Sr}$ must be
%\begin{eqnarray}
$\min \Delta\mu_{\rm Sr} = \Delta E_f^{\rm SrTiO_3} 
-\max {\Delta\mu_{\rm Ti}}
-3\Delta\mu_{\rm O}$
%,\label{eq:Sr-min-3}\\
%\end{eqnarray}
according the equilibrium condition for SrTiO$_3$. Taking into account that the formation of TiO$_2$ must be unfavorable
%\begin{eqnarray}
(that is, $ \max \Delta\mu_{\rm Ti}= \Delta E_f^{\rm TiO_2} 
        -2\Delta\mu_{\rm O}$)
%\end{eqnarray}
we then have
\begin{equation}
\min \Delta\mu_{\rm Sr} = 
    \Delta E_f^{\rm SrTiO_3} -\Delta E_f^{\rm TiO_2} -\Delta\mu_{\rm O}.\label{eq:Sr-min}
\end{equation} 
The maximum value of $\Delta\mu_{\rm Sr}$, on the other hand, is determined the condition of unfavourable formation of SrO,
\begin{eqnarray}
\max \Delta\mu_{\rm Sr}=  \Delta E_f^{\rm SrO} -\Delta\mu_{\rm O}.\label{eq:Sr-max}
\end{eqnarray}

Regarding $\Delta\mu_{\rm O}$ itself, this potential should be high enough so that LaNiO$_3$ does not decompose into LaNiO$_2$. This implies
$\Delta E_f^{\rm LaNiO_3} < \Delta E_f^{\rm LaNiO_2} + \Delta\mu_{\rm O}$ 
which, according to the energies listed in Table~\ref{tab:bulks}, further means that $-2.50$~eV~$ <\Delta\mu_{\rm O}$. This is marked by the dashed yellow line in Fig. \ref{fig:LaNiOnSTO}. 
At the same time, LaNiO$_3$ should also be stable with respect its decomposition into La$_2$O$_3$ and NiO. This, however, is ensured by the fact that the inequality 
$
\Delta E_f^{\rm LaNiO_3} < \frac{1}{2} \Delta E_f^{\rm La2O_3} +\Delta E_f^{\rm NiO} +\frac{1}{2}\Delta\mu_{\rm O} 
$
holds whenever $-2.50$~eV~$ <\Delta\mu_{\rm O}$.  

\begin{table}[t!]
\begin{tabular}{l c c c c r}
\hline \hline
            & $\Delta E_f$ (eV) & $E_{\rm tot}$  (eV) &   $a$ (\AA)  &  $c$ (\AA ) &   Space group   \\
            \hline
LaNiO$_3$   & $-$12.27 & $-$35.90 &  3.829       &  3.829 &        $P4/mmm$        \\            
LaNiO$_2$   & $-$9.77  & $-$28.96 &  3.936       &  3.393 &        $P4/mmm$       \\
LaNiO       & $-$5.42  & $-$20.18 &  3.584       &  3.842 &        $P4/mmm$       \\
La$_2$O$_3$ & $-$18.78 & $-$41.84 &  3.937       &  6.181 &       $P6_3/mmc$       \\
NiO         & $-$1.48  & $-$11.36 &  4.151       &  4.151 &        $Fm\bar{3}m$        \\
La          & -        &  $-$4.88 &  3.767       & 12.129 &        $P6_3/mmc$       \\
Ni          & -        &  $-$5.44 &  3.499       &  3.499 &        $Fm\bar{3}m$        \\ \hline 
SrTiO$_3$   & $-$17.35 & $-$40.11 &  3.940       &  3.940 &         $Pm\bar{3}m$       \\
SrO         & $-$5.99  & $-$12.06 &  5.203       &  5.203 &          $Fm\bar{3}m$      \\
TiO$_2$     & $-$9.17  & $-$25.87 &  4.160       &  8.579 &        $I4_1/amd$      \\
Sr          & -        &  $-$1.64 &  6.038       &  6.038 &          $Fm\bar{3}m$      \\
Ti          & -        &  $-$7.84 &  2.939       &  4.640 &        $P6_3/mmc$        \\  \hline
CaH$_2$O$_2$& $-$10.55 & $-$28.10 &  3.618       &  4.913 &       $P\bar{3}m1$ 
   \\
CaH$_2$     & $-$1.69  & $-$10.38 &  * &  6.760 &       $Pnma$ 
\\
\hline \hline
\end{tabular}
\caption{
Formation energy and structural parameters of the different compounds used to determine the chemical potentials that define the interface energy according to Eq. \eqref{Einterface}.
The formation energy of LaNiO$_2$, for example, is defined as $\Delta E_f^{\text{LaNiO}_2} = E_\text{tot}^{\text{LaNiO}_2} - E_\text{tot}^{\text{La}} -E_\text{tot}^{\text{Ni}}-E_\text{tot}^{\text{O}_2}  $, and a similar definition applies for the rest of compounds. *In the case of the orthorombic CaH$_2$ the lattice parametters perpendicular to the $c$ axis are $a= 5.892$~{\AA} and $b=3.568$~{\AA}.}
\label{tab:bulks}
\end{table}

We then note that Eq.~\eqref{eq:La-max-3} must be used with caution since, according to Table~\ref{tab:bulks}, one obtains a positive value of $\max \Delta\mu_{\rm La}$ for $\Delta\mu_{\rm O}>-1.92$ eV. This is incompatible with the fact that the chemical potentials are such that $\Delta\mu_{\rm X}<0$. Consequently, in the O-rich situation in which $\Delta\mu_{\rm O}>-1.92$ eV we can safely put $\max \Delta\mu_{\rm La}=0$.

Thus, subtracting Eq.~\eqref{eq:Sr-max} from Eq.~\eqref{eq:La-min-3} we obtain the minimum value for the difference $\Delta\mu_{\rm La}-\Delta\mu_{\rm Sr}$ as
\begin{align}
\min(\Delta\mu_{\rm La}-\Delta\mu_{\rm Sr}) =
  \Delta E_f^{\rm LaNiO_3} - \Delta E_f^{\rm SrO} -2\Delta\mu_{\rm O},
\label{minLa-Sr}
\end{align}
while subtracting Eq.~\eqref{eq:Sr-min} from Eq.~\eqref{eq:La-max-3} we obtain its maximum value
%\begin{widetext}
\begin{align}
\max (\Delta\mu_{\rm La}-\Delta\mu_{\rm Sr} )
  &
  =\begin{cases}
  \Delta E_f^{\rm LaNiO_3} - \Delta E_f^{\rm SrTiO_3}
  + \Delta E_f^{\rm TiO_2}-2\Delta\mu_{\rm O} & \text{ ($\Delta\mu_{\rm O}<-1.92$ eV, O-poor)},\\
  \frac{1}{2}\Delta E_f^{\rm La_2O_3} - \Delta E_f^{\rm SrTiO_3} 
  + \Delta E_f^{\rm TiO_2}-\frac{1}{2}\Delta\mu_{\rm O} & \text{ ($\Delta\mu_{\rm O}>-1.92$ eV, O-rich)}.
  \end{cases}
\label{maxLa-Sr}\end{align}
%\end{widetext}
Eqs. \eqref{minLa-Sr} and \eqref{maxLa-Sr} give rise to the yellow region in Fig. \ref{fig:LaNiOnSTO}, which is determined according to the values listed in Table~\ref{tab:bulks}. 
We find, in particular, 
\begin{align}
       -1.28 \text{ eV}<\Delta\mu_{\rm La} &-\Delta\mu_{\rm Sr}<+0.91 
       \text{ ($\Delta\mu_{\rm O}=-2.5$ eV)}, 
       \\
        -6.28  \text{ eV}<\Delta\mu_{\rm La} &-\Delta\mu_{\rm Sr} <-1.21 \text{ ($\Delta\mu_{\rm O}=0$ eV)}.
\end{align}

\subsection{Topotactic reduction}

In \cite{li-nature19} the LaNiO$_3$/SrTiO$_3$ sample was vacuum-sealed together with ---but not in direct contact to--- CaH$_2$ powder for the topotactic reaction. Then a gas-phase reaction is activated upon heating in which oxygen is removed from LaNiO$_3$ and incorporated the powder yielding CaH$_2$O$_2$. In this reaction $\Delta\mu_{\rm O}$ is lowered so that LaNiO$_3$ is forced to loose oxygen atoms (i.e. below $-2.5$~eV). 
Its eventual value have to be such that
\begin{equation}
    \Delta E_f^{\rm CaH_2O_2} < \Delta E_f^{\rm CaH_2} +2\Delta\mu_{\rm O}.  
\end{equation}
According to Table~\ref{tab:bulks} that means $\Delta\mu_{\rm O}>-4.43$~eV, which is marked by the dashed blue line in Fig. \ref{fig:LaNiOnSTO}.
We note that $-4.43$~eV is just slightly below the minimum value of $\Delta\mu_{\rm O}$ in LaNiO$_2$ (see below), which means that a more reducing environment would produce LaNiO instead of LaNiO$_2$ out of LaNiO$_3$.

Regarding the possible reduction of SrTiO$_3$ to SrTiO$_2$, for this to happen the O chemical potential has to satisfy the condition
$\Delta E_f^{\rm SrTiO_3} > \Delta E_f^{\rm SrTiO_2} + \Delta\mu_{\rm O}$ 
which, according to the energies listed in Table~\ref{tab:bulks}, further means that $\Delta\mu_{\rm O} < - 5.65$~eV.
This, however, is incompatible with the limit set by the topotactic reduction reaction ($- 4.4$ eV $<\Delta\mu_{\rm O}$). Thus, we conclude that the topotactic reduction can certainly remove O from LaNiO$_3$ but will not be able to transform SrTiO$_3$ into SrTiO$_2$. 

\subsection{LaNiO$_2$/SrTiO$_3$ heterostructure}

The energy difference between the La and LaO interfacial configurations reads
\begin{align}
    \Delta E_\text{interface}^{\rm La - LaO}
    %&= \frac{1}{2} \big( E_{\rm tot}^{\rm (La)}-E_{\rm tot}^{\rm (LaO)}\big)+\mu_{\rm O} \nonumber \\
    &=\frac{1}{2} \big( E_{\rm tot}^{\rm (La)}-E_{\rm tot}^{\rm (LaO)}+E_{\rm tot}^{\rm O_2}\big)+\Delta\mu_{\rm O}. \label{LaO-Sr}    
\end{align}
Accordingly, the LaO interface will be favored over the La one if $\Delta\mu_{\rm O}> -4.38$ eV (and vice versa). 
Similarly, we find that the SrO interface will be favored over the Sr one if $\Delta\mu_{\rm O}> -3.67$ eV. The boundaries defined by these relative stabilites are indicated by the thick vertical lines in blue in Fig. \ref{fig:LaNiOnSTO}, and apply for both the topotactic reduction and the direct growth of the LaNiO$_2$/SrTiO$_3$ heterostructure. 

Consider now the formation-energy difference between the SrO and LaO, Sr and LaO, and Sr and La interfaces:
\begin{align}
    \Delta E_\text{interface}^{\rm SrO - LaO}
    &=\frac{1}{2} \left( E_{\rm tot}^{\rm (SrO)}-E_{\rm tot}^{\rm (LaO)}\right)+E_{\rm tot}^{\rm La}-E_{\rm tot}^{\rm Sr}
    %\nonumber \\ &\quad 
    +\Delta\mu_{\rm La}-\Delta\mu_{\rm Sr}, \label{SrO-LaO} 
    \\
    \Delta E_\text{interface}^{\rm Sr - LaO}
    &=\frac{1}{2} \left( E_{\rm tot}^{\rm (Sr)}-E_{\rm tot}^{\rm (LaO)}+E_{\rm tot}^{\rm O_2}\right)+E_{\rm tot}^{\rm La}-E_{\rm tot}^{\rm Sr}
    %\nonumber \\ &\quad 
    +\Delta\mu_{\rm La}-\Delta\mu_{\rm Sr}+\Delta\mu_{\rm O}. \label{Sr-LaO}    
    \\
    \Delta E_\text{interface}^{\rm Sr - La}
    &=\frac{1}{2} \left( E_{\rm tot}^{\rm (Sr)}-E_{\rm tot}^{\rm (La)}\right)+E_{\rm tot}^{\rm La}-E_{\rm tot}^{\rm Sr}
    %\nonumber \\ &\quad 
    +\Delta\mu_{\rm La}-\Delta\mu_{\rm Sr}, \label{Sr-La} 
\end{align}
According to the computed values of the total energies (Tables~\ref{tab:SL-struct} and \ref{tab:bulks}), we find that the LaO interface is energetically favored with respect to the SrO one if $\Delta\mu_{\rm La}-\Delta\mu_{\rm Sr}>-1.54$ eV. 
The LaO configuration is additionally favored over the Sr one if $\Delta\mu_{\rm La}-\Delta\mu_{\rm Sr}+\Delta\mu_{\rm O}>-5.21$ eV.
Finally, we find that the La interface is energetically favored with respect to the Sr one if $\Delta\mu_{\rm La}-\Delta\mu_{\rm Sr}>-0.83$ eV. 
These considerations are relevant for the direct growth of the LaNiO$_2$/SrTiO$_3$ heterostructure only (since they imply La~$\leftrightarrow$~Sr replacements that, in principle, cannot be obtained by means of the topotactic reduction of the perovskite), and are indicated by thin blue lines in Fig. \ref{fig:LaNiOnSTO}.

In terms of $\Delta\mu_{\rm O}$, the realization of the nickelate in its infinite-layer phase requires the following. On one hand, LaNiO$_2$ has to be more stable than LaNiO$_3$.
That is, $
   \Delta E_f^{\rm LaNiO_2} <  \Delta E_f^{\rm LaNiO_3} - \Delta\mu_{\rm O}$.
At the same time, LaNiO$_2$ must be more stable than LaNiO + O. That is,
$
   \Delta E_f^{\rm LaNiO_2} <  E_f^{\rm LaNiO} + \Delta\mu_{\rm O} 
$. Taken together, these two conditions tell us that 
\begin{align}
%\underbrace{
-4.35~\text{eV}
%}_{\Delta E_f^{\rm LaNiO_2} -\Delta E_f^{\rm LaNiO}}
<\Delta\mu_{\rm O} < 
%\underbrace{
-2.50~\text{eV}
%}_{\Delta E_f^{\rm LaNiO_3} -\Delta E_f^{\rm LaNiO_2}}
\label{muOrange}
\end{align}
according to the formation energies listed in Table~\ref{tab:bulks}.
The limits of this interval correspond to the dashed lines in Fig. \ref{fig:LaNiOnSTO}. 
We note that NiO is never stable within such a $\Delta\mu_{\rm O}$ interval, so that $\Delta\mu_{\rm Ni}$ can reach 0 ({\it i.e.} ${\max \Delta\mu}_{\rm Ni} =0$). Further, 
the potentials need to be such that
\begin{eqnarray}
        \min \Delta\mu_{\rm La} &=& \Delta E_f^{\rm LaNiO_2} 
        %- \cancelto{0}{\max \Delta\mu}_{\rm Ni}
        -2\Delta\mu_{\rm O},\\
        \max \Delta\mu_{\rm Sr} &=& \Delta E_f^{\rm SrO} - \Delta\mu_{\rm O},
\end{eqnarray}
By subtracting these equations and using Table~\ref{tab:bulks} we find
that 
\begin{align}
       \min [\Delta\mu_{\rm La} -\Delta\mu_{\rm Sr}]+ \Delta\mu_{\rm O} = 
       %\underbrace{
       -3.78~\text{eV}
       %}_{\Delta E_f^{\rm LaNiO_2} - \Delta E_f^{\rm SrO}}
       .
\label{eq:eq} \end{align}
This further defines the blue region in Fig. \ref{fig:LaNiOnSTO}.

\section{Electronic band structure within LDA + $U$ \label{a:LDA+U}}
\begin{figure*}[h!]
%LDA + $U$\\
%\vspace{4pt}
\includegraphics[height=.2125\textwidth]{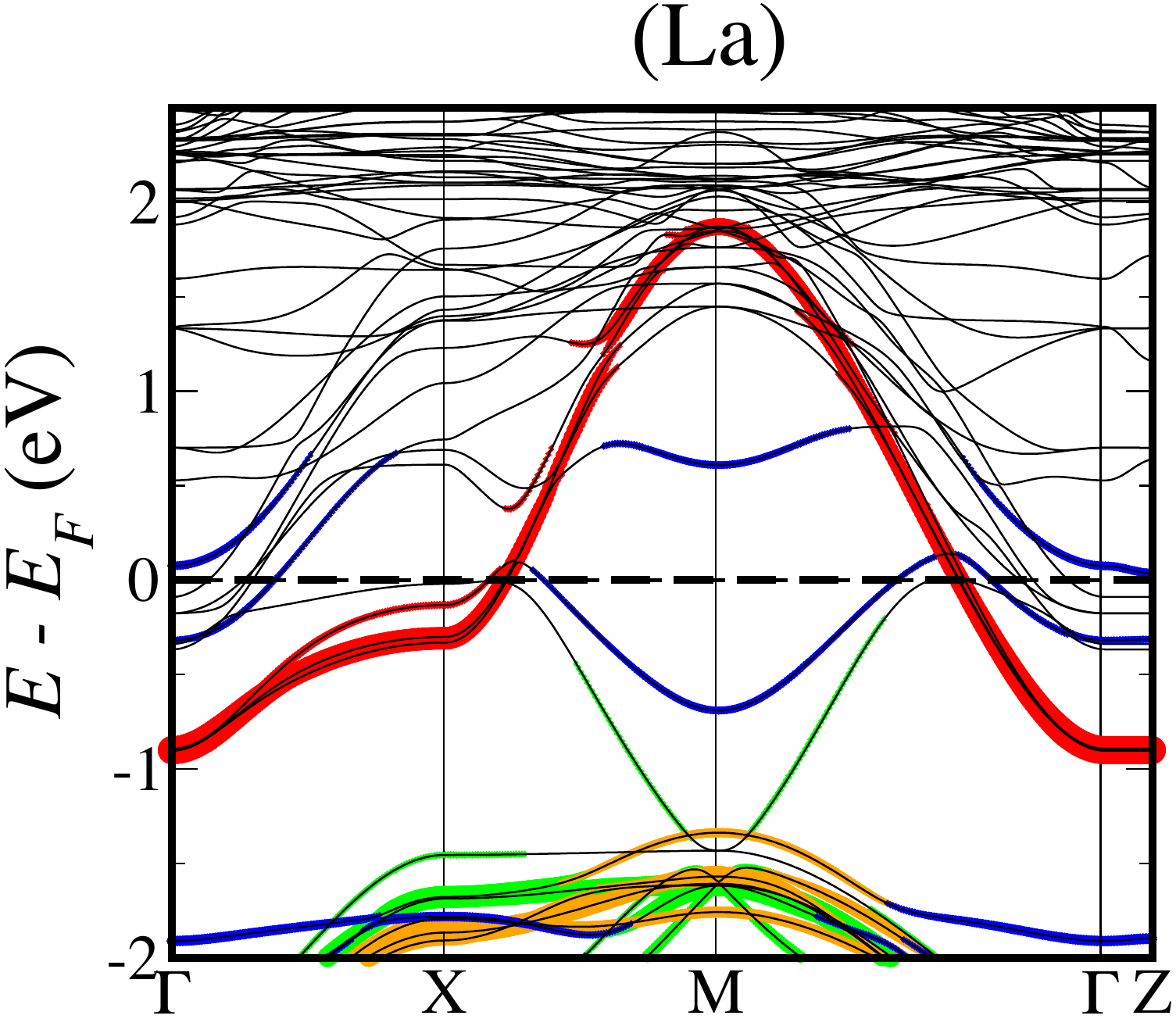}
\includegraphics[height=.2125\textwidth]{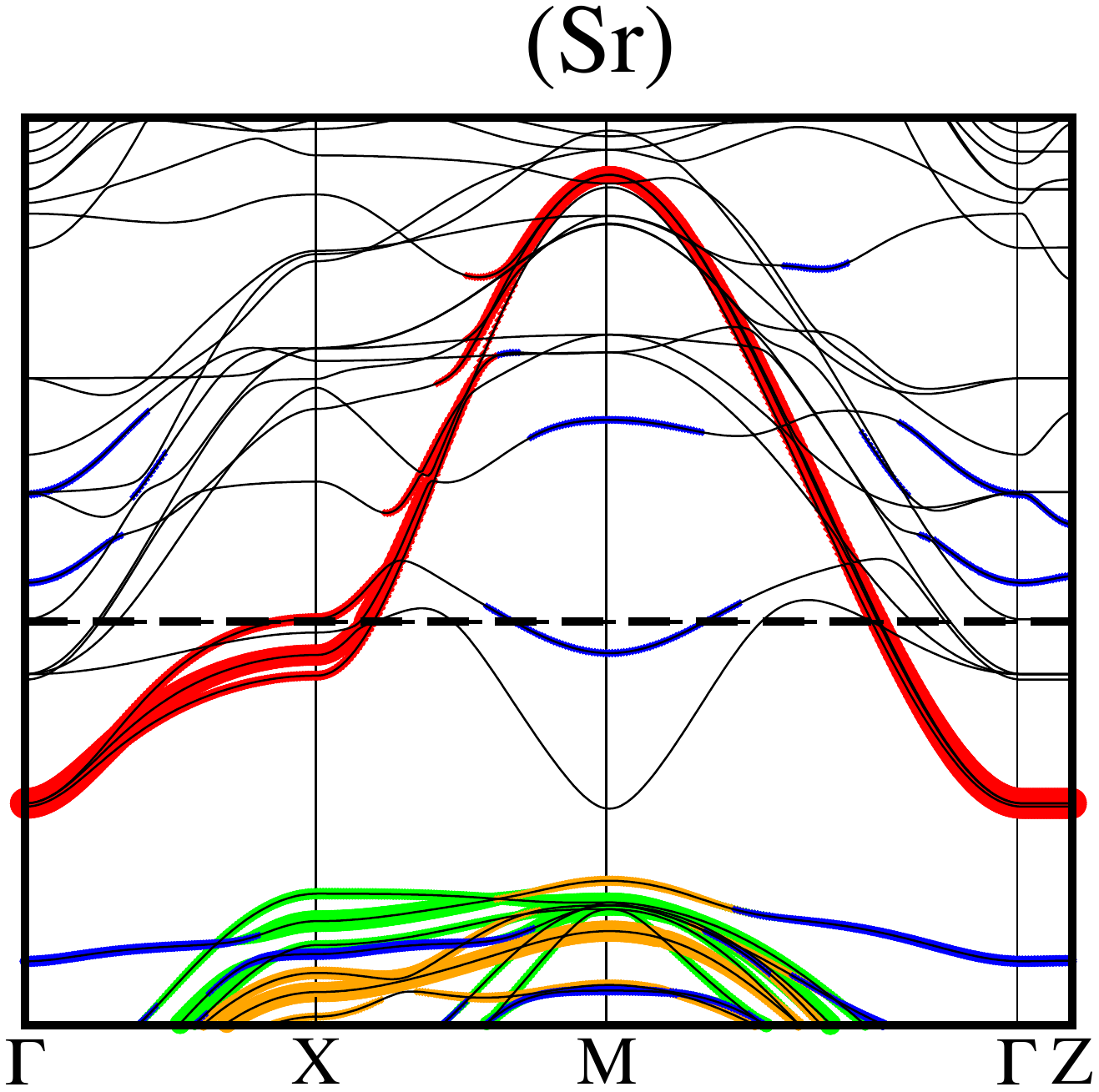}
\includegraphics[height=.2125\textwidth]{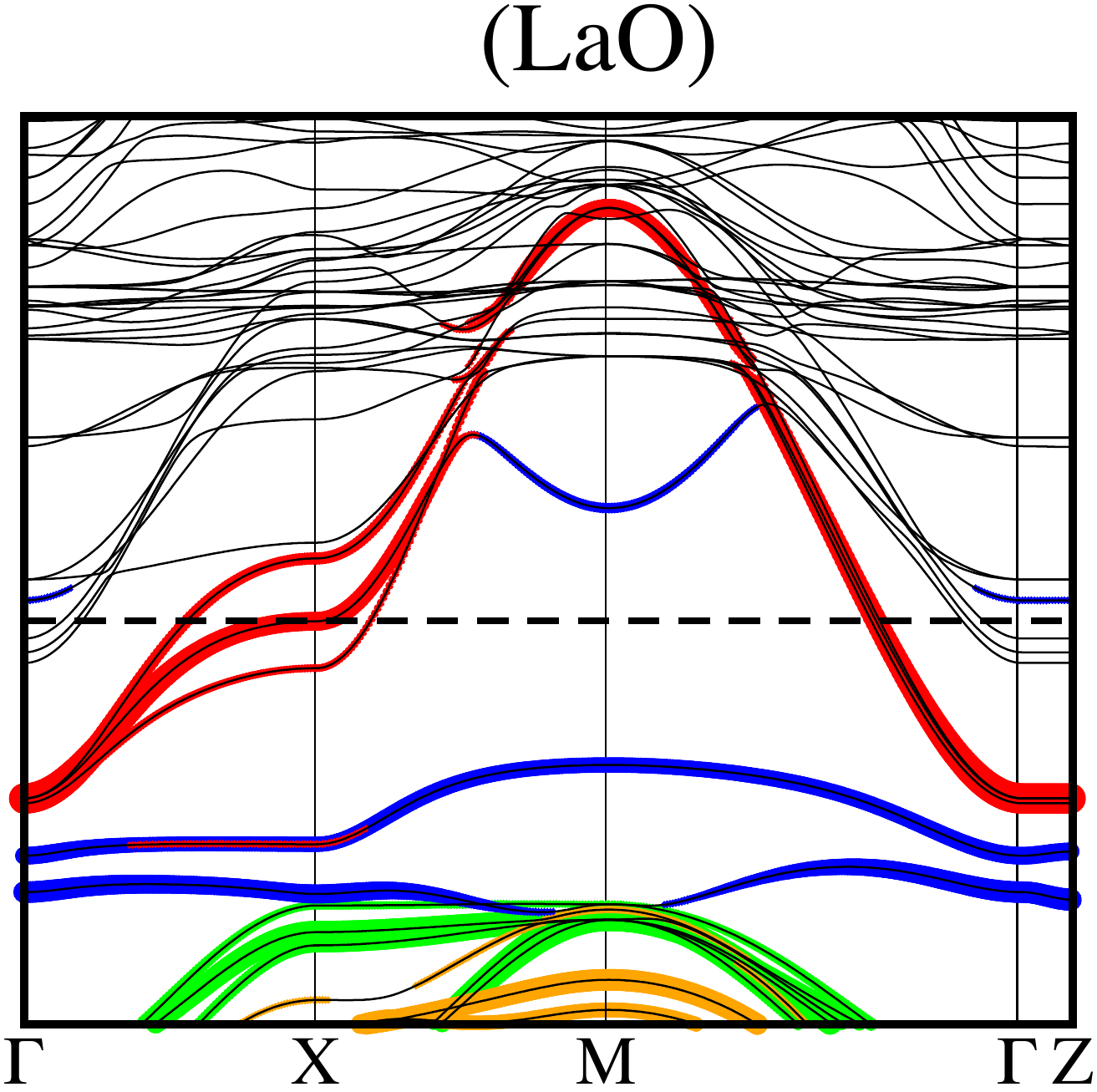}
\includegraphics[height=.2125\textwidth]{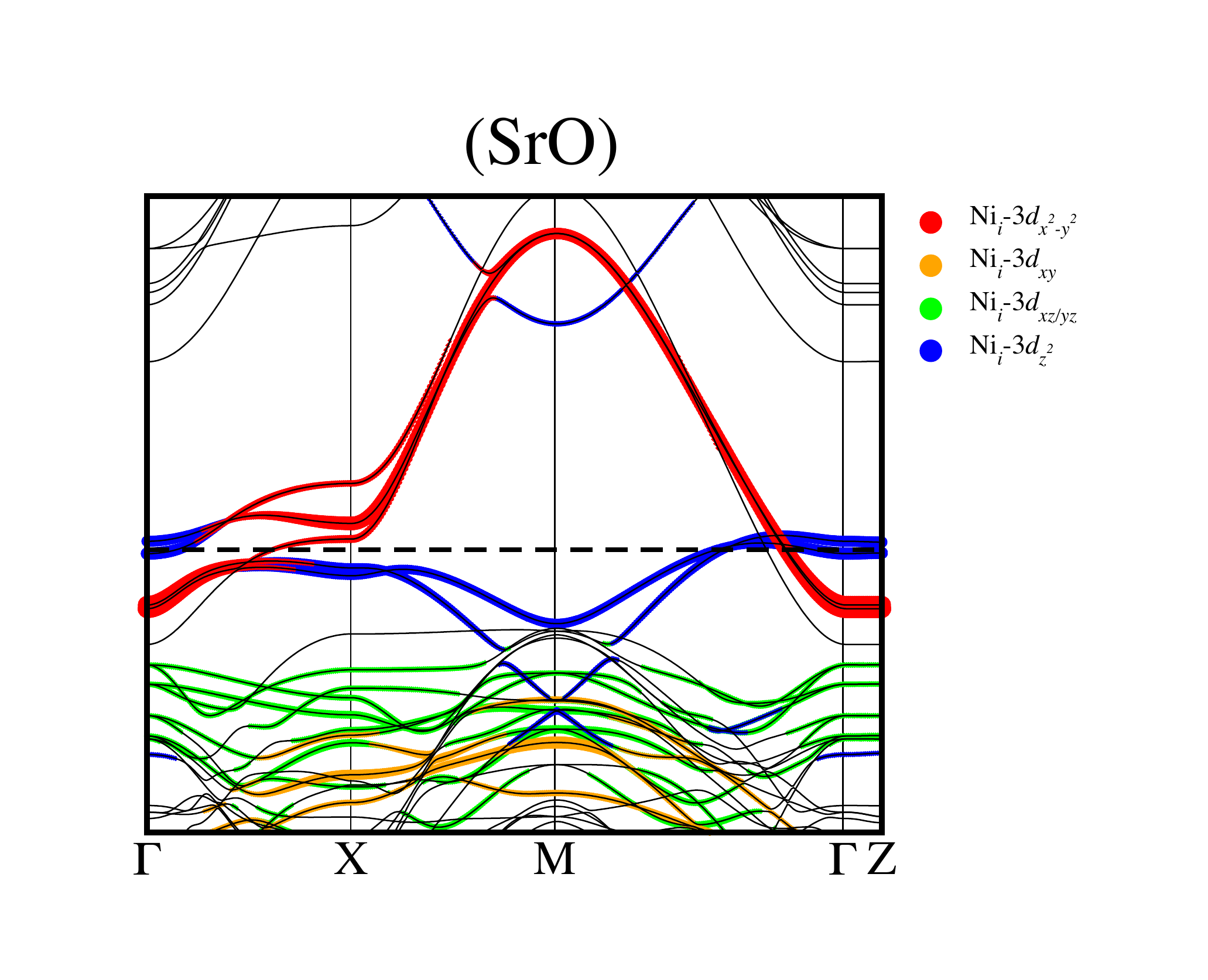}
  \caption{Band structure of the LaNiO$_2$/SrTiO$_3$ symmetric supercells for different interfacial configurations in the LDA + $U$ approximation ($U_\text{Ni} = U_\text{Ti} = 4$~eV). The line thickness and the different colors highlight the interfacial contributions according to their atomic-orbital content.
  }  
\label{fig:lda+u}
\end{figure*}

\blue{
The band structure calculations reported in the main text were performed with the LDA exchange and correlation functional as explained in Sec. \ref{methods}. This choice is motivated, in particular, by the fact that LDA still reproduces remarkably well the low-energy features obtained for bulk LaNiO$_2$ when the missing correlations beyond this approximation are included at the GW level \cite{olevano20}. In the following we take into account these correlations within the LDA + $U$ scheme and show that interfacial band structure of the LaNiO$_2$/SrTiO$_3$ heterostructure remains qualitatively the same near the Fermi level. This provides additional support to our results. 

Specifically, we take into account intra-atomic correlations for both Ni and Ti atoms using the LDA + $U$ method in the ‘fully localized limit’ (FLL). Similarly to previous works, we consider $U_\text{Ni} = U_\text{Ti} = 4$~eV \cite{botana19,Zhong_2008}. Fig.~\ref{fig:lda+u} shows the resulting band structure as a function of the interfacial atomic configuration in the paramagnetic case (see Fig. \ref{fig:interfaces}). 

Compared to the LDA results (see Fig. \ref{fig:SLbands}), all the interfacial Ni-3$d$ bands but the Ni-3$d_{x^2-y^2}$ ones are shifted down with respect to the Fermi level irrespective of the interfacial configuration. This is totally in tune with the effect that the Hubbard term has in bulk LaNiO$_2$. At the same time, the position of the Ni-3$d_{x^2-y^2}$ bands relative to the Fermi level as well as their dispersion remain essentially unchanged. For the SrO configuration, in particular, there is a slight modification of the $d_{x^2-y^2}$-$d_{z^2}$ crossings that, however, still take place at the Fermi level. In any case, the interfacial Ni-3$d_{z^2}$ bands remain remarkably flat (also for the LaO case). Furthermore, compared to the bulk, both LDA and LDA + $U$ disclose interfacial changes in the self-doping effect due to the electron pockets at $\Gamma$ that are totally alike.
This confirms the robustness of our results.
}

\end{document}